\documentstyle[aps,multicol,prl,epsf]{revtex}

\begin{document}

\draft

\title{CrO$_2$: a self-doped double exchange ferromagnet}

\author{M.A. Korotin and V.I.  Anisimov}
 
\address{Institute of Metal Physics, Ural Division of Russian Academy
of Sciencies, \\ 620219 Ekaterinburg GSP-170, Russia}

\author{D.I.  Khomskii and G.A.  Sawatzky}

\address{ Solid State Physics Dept of the Materials Science Center,
University of Groningen, \\ Nijenborgh 4, 9747 AG Groningen, The
Netherlands}

\maketitle

\begin{abstract}

Band structure calculations of CrO$_2$ carried out in the LSDA+U
approach reveal a clear picture of the physics behind the metallic
ferromagnetic properties.  Arguments are presented that the metallic
ferromagnetic oxide CrO$_2$ belongs to a class of materials in which
magnetic ordering exists due to double exchange (in this respect
CrO$_2$ turns out to be similar to the CMR manganates).  It is
concluded that CrO$_2$ has a small or even negative charge transfer
gap which can result in self-doping.  Certain experiments to check the
proposed picture are suggested.

\end{abstract}

\pacs{75.50.Ss, 71.20.-b, 75.30.Et} 
 
% 75.50.Ss - Magnetic recording materials
% 71.20.-b - Electron DOS and band structure of crystalline solids 
% 75.30.Et - Exchange and superexchange interactions in magnetic 
%            materials

\begin{multicols}{2}

There has been a revival in interest in $3d$ transition metal oxides
during the last decade.  This was initially stimulated by the
discovery of High-T$_c$ superconductivity in complex copper oxides,
and more recently by the active study of the colossal
magnetoresistance manganates (CMR) La$_{1-x}$(Ca,Sr)$_x$MnO$_3$.
These latter systems in the most interesting composition range are
metallic ferromagnets.  This is interesting in itself because
ferromagnetic ordering is rare among the oxides:  most of them are
antiferromagnetic or ferrimagnetic with dominating antiferromagnetic
interactions.  We need to meet certain special conditions to stabilize
ferromagnetism.  One of the main mechanisms invoked to explain
ferromagnetic ordering in these systems is the double exchange
mechanism \cite{Zener}, although it is not the only one
\cite{KhS_rev}.

Another very well known ferromagnetic and metallic compound is
chromium dioxide CrO$_2$ widely used in magnetic recording tapes.  In
its formal $4+$ valence state Cr has two $3d$ electrons in t$_{2g}$
orbitals which in a simple picture of strong correlations would
suggest a Mott insulating-like ground state with S=1 local moments and
most likely antiferromagnetic spin order.  This seems to be about as
far from the actual observed properties as one can get.  In this
letter we will address this problem using band structure methods
supplemented with local Coulomb and exchange interactions (LSDA+U,
\cite{LSDA+U}) interpreted in terms of local electronic
configurations.  We will argue that the $d$ electrons can be divided
into a localized "core" of spin~$\frac{1}{2}$ and an itinerant $d$
electron propagating through these "cores" resulting in a double
exchange-like mechanism for the ferromagnetic order much as in the
manganates.  We also show that strong electron correlation effects do
not in this case lead to an insulating ground state for reasonable
values of the $d$-$d$ Coulomb interaction and perfect stoichiometry
because CrO$_2$ should be viewed as a small or even negative charge
transfer gap \cite{neggap} material in the Zaanen-Sawatzky-Allen (ZSA)
scheme \cite{ZSA} quite unlike the parent CMR material LaMnO$_3$ which
is a Mott-Hubbard insulator.  This leads for CrO$_2$ quite naturally
to a phenomenon which could be referred to as self-doping resulting in
a non-integral $3d$ band occupation \cite{LJP}.

CrO$_2$ is a ferromagnetic metal with a saturation magnetic moment of
2.00~$\mu_B$ and a low temperature resistivity with a nearly T$^2$
temperature dependence \cite{Chamb_rev}.  Band structure calculations
in LSDA \cite{Schwarz} explain this behaviour as that of a
half-metallic ferromagnet with a gap in the minority spin band
resulting in the integral magnetic moment per formula unit in spite of
the strong covalency effects.  However the magnetic susceptibility in
the paramagnetic phase exhibits a Curie-Weiss-like behaviour with a
local moment of also 2~$\mu_B$ indicating that the mechanism for
ferromagnetic behaviour is not a band (Stoner-like) mechanism.  Local
moments indicate strong correlation effects and this was in fact
suggested by early photoemission data \cite{Kamper} which looked more
like those of a semiconductor with a vanishing density of states (DOS)
at the Fermi energy and a spin polarization of nearly 100~\% for
binding energies about 2~eV below the Fermi level.  To resolve this
controversy we carried out the calculations in the LSDA+U approach
which is superior to LSDA since it can indeed yield a gap if the local
Coulomb interactions are large enough.

\begin{figure}
\epsfxsize=8.4cm
\vspace{1mm}
\centerline{ \epsffile{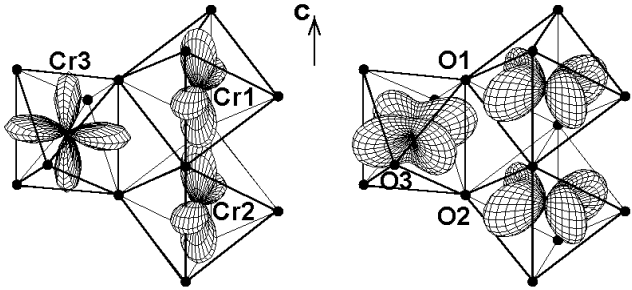} } 
\vspace{1mm}
\narrowtext

\caption {Angular distribution of $xy$ (left) and ($yz+zx$) (right)
electron spin density for the nearest Cr neighbors.  Direction of
crystallographic $c$-axis is shown by arrow.  Solid circles denote
oxygen atoms.}

\label{orb}
\end{figure}

CrO$_2$ has a rutile structure (space group D$_{4h}^{14}$:
P4$_2$/$mnm$) in which the unit cell consists of two formula units.
The Bravais lattice is tetragonal ($c$/$a$=0.65958) with a lattice
constant of $a$=4.421~\AA \cite{cryststr}.  The Cr atoms form a
body-centered tetragonal lattice and are surrounded by distorted
oxygen octahedra.  The octahedra surrounding Cr at the body's center
and corner positions differ by a 90$^{\circ}$ rotation about the $c$
axis (see also Fig.\ref{orb} where Cr1 and Cr2 atoms are in the
corners of the unit cell and Cr3 atom is in the body center position).

%%%%%%%%%%%%%%%%%%%%%%%%%%%%%%%%%%%%%%%%%%%%%%%%%%%%%%%%%%%%%
%%%%%%%%%%%%%%%%%%%%%%%%%%%%%%%%%%%%%%%%%%%%%%%%%%%%%%%%%%%%%

The simplest CrO$_6$ cluster calculations showed, that this kind of
distortion of the octahedra (elongation along $c$ axis) leads to the
new natural basis for the t$_{2g}$ orbitals:  $xy$, $(yz+zx)$ and
$(yz-zx)$ in a local coordinate system (LCS) for every octahedron.  In
this LCS the $z$-axis is directed to the apex oxygen and the $x$ and
$y$ axes -- to the basal plane oxygens.  Again refering to the
Fig.\ref{orb}, in which for Cr3 atom z-axis is directed to the O3
oxygen and $x$, $y$ axes -- to the O1 and O2 oxygens.  At the same
time, the O1 and O2 become apex oxygens for Cr1 and Cr2, and $z$ axis
is directed to these oxygens there.  We will demonstrate below that
this distorted structure is responsible for rather peculiar properties
of the $d$ bands.

LSDA+U calculations were performed in the linearized muffin-tin
orbitals (LMTO) approach \cite{lmto} with atomic spheres radii of
2.06~a.u.  for Cr and O and 1.78 and 1.62~a.u.  for the two kinds of
empty spheres used (8 per unit cell in total) to fill the empty space
as much as possible.  The spherical harmonics were expanded to the
value of l$_{max}$=3 and 2 for the atomic and empty spheres,
respectively.  For the Brillouin zone (BZ) integration 1300~{\bf
k}-points were used.  The screened U and J parameters used in the
LSDA+U scheme were calculated by constrain method taking into account
participation of e$_g$ electrons in screening of the t$_{2g}$
electrons \cite{calcU} and found to be 3 and 0.87~eV, respectively.

\begin{figure} 
\epsfxsize=16cm 
\centerline{ \epsffile{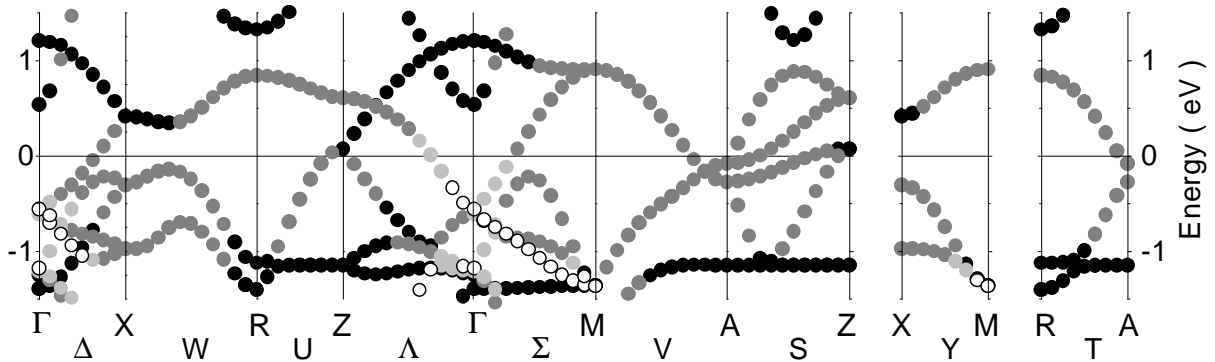} }
\vspace{1mm}

\caption {LSDA+U band structure of CrO$_2$ for majority
spin-sublattice in the vicinity of the Fermi level.  Open circles
denote the contribution of $d$ states less than 25~\%, light gray
circles -- between 25 and 50~\%, gray -- between 50 and 75~\%, and
black -- predominantly $d$ character of the band.  Fermi level is
denoted by horizontal solid line.}

\label{bands}
\end{figure}

\begin{figure}
\epsfxsize=8.5cm
\centerline{ \epsffile{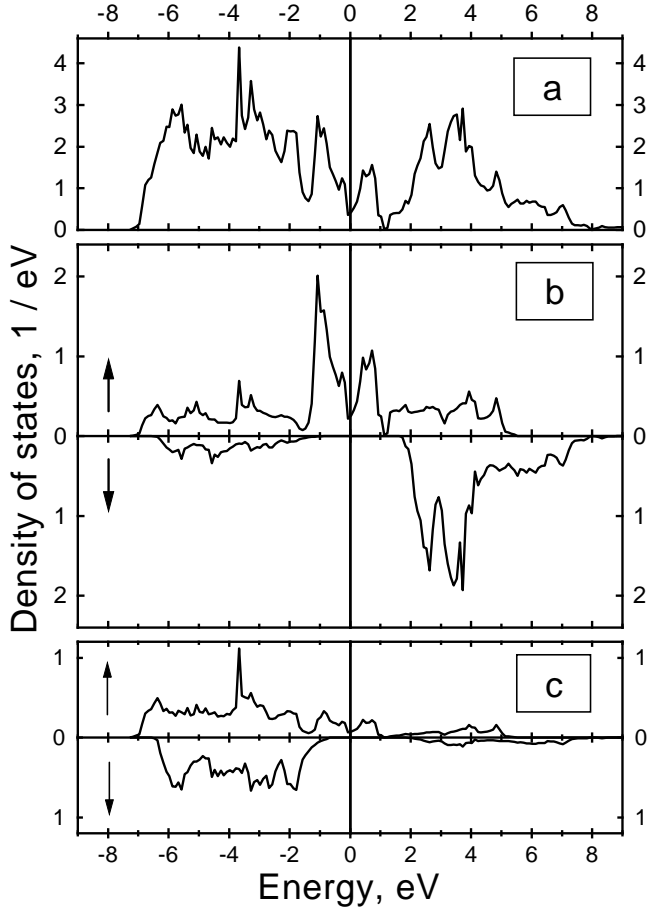} }
\narrowtext
\vspace{1mm}

\caption {Total (per formula unit and both spins, a) and partial Cr
$3d$- (b) and O $2p$- (c) density of states of CrO$_2$.}

\label{dos} 
\end{figure}

\begin{figure}
\epsfxsize=6.5cm
\centerline{ \epsffile{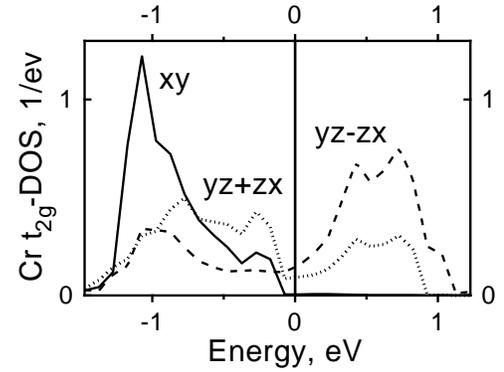} }
\vspace{1mm}
\narrowtext

\caption {Partial Cr t$_{2g}$ majority spin density of states in the
vicinity of the Fermi level.}

\label{pdos}
\end{figure}

The results of this calculation are shown in Figs.
\ref{bands}-\ref{pdos}.  First, for U=3~eV the material is found to be
a half-metallic ferromagnet, similar to the previous LSDA results
\cite{Schwarz}.  However, inclusion of the electron correlations
modifies the electronic structure.  Together with the shift of the
minority spin DOS from the bottom of the conduction band to higher
energy, there appears a dip in the majority spin DOS at the Fermi
level which can be taken as an indication of the tendency toward a gap
formation.  The decrease of the DOS at E$_F$ in comparison with that
obtained in \cite{Schwarz} may be partially responsible for the
reduced signal close to E$_F$ observed in photo-electron spectroscopy
\cite{Kamper}.  In our calculations, however, CrO$_2$ is still a metal
albeit with the reduced DOS at the Fermi level.  An explanation of
this discrepancy may lie, e.g., in the modified situation at the
surface.  Note also that more recent photoemission data
\cite{Fujimori} show a finite DOS at E$_F$, and the spectrum obtained
in \cite{Fujimori} is in rather good agreement with our calculations.

The tendency to open a gap in an energy spectrum which we noticed
above was confirmed by calculation with larger values of U.  It was
found that the gap would indeed open at U$\geq$6~eV which would make
CrO$_2$ a Mott-Hubbard insulator.

There are other aspects of the calculation that are very interesting
and may provide the clues to the basic mechanism for the ferromagnetic
metal behaviour.  First notice the almost dispersionless majority spin
band at about 1~eV below E$_F$ in Fig.\ref{bands} over a large region
of the BZ which as is indicated by the black circles has almost pure
$d$ character.  This is observed as a peak in the DOS in
Figs.\ref{dos} and \ref{pdos}, and corresponds to strongly localized
$xy$ orbitals completely occupied by one majority spin electron.
Because these states have almost pure $d$ character they are very
sensitive to the size of U and shift further away from E$_F$ as U
increases.

%%%%%%%%%%%%%%%%%%%%%%%%%%%%%%%%%%%%%%%%%%%%%%%%%%%%%%%%%%%%%
\end{multicols}
\widetext

\begin{multicols}{2}
%%%%%%%%%%%%%%%%%%%%%%%%%%%%%%%%%%%%%%%%%%%%%%%%%%%%%%%%%%%%%

%%%%%%%%%%%%%%%%%%%%%%%%%%%%%%%%%%%%%%%%%%%%%%%%%%%%%%%%%%%%%
%%%%%%%%%%%%%%%%%%%%%%%%%%%%%%%%%%%%%%%%%%%%%%%%%%%%%%%%%%%%%

The other t$_{2g}$ $d$-bands are strongly hybridized with O and form
dispersive bands crossing the Fermi level.  One can see from
Fig.\ref{pdos} that predominantly the ($yz+zx$) is occupied at each
site, see also Fig.\ref{orb}.  An admixture of oxygen states strongly
softens the influence of U on these bands and that is the reason that
such a high value of U=6~eV is required for a gap to open in these
bands.  This indicates that we are dealing with a system of localized
S=$\frac{1}{2}$ $xy$ electrons exchange coupled to broader band-like
and strongly hybridized ($yz+zx$) states.  In our case the bandwidth
of $d$-$p$ ${\pi}$-bands is comparable with the Hund's rule
intraatomic exchange ($\sim$1~eV).  This is reminiscent of a double
exchange ferromagnetic mechanism of Zener \cite{Zener}.  This is also
consistent with the half-metallic nature of this material.  Thus, the
(half-)metallic character and ferromagnetism are indeed closely
connected.

An extra confirmation of this conclusion came from the calculation, in
which we artificially imposed an antiferromagnetic ordering.  The
simplest two-sublattice antiferromagnetic structure was chosen, with
the spin in the body center sites opposite to the spins at the corners
of the unit cell.  The band structure obtained has a small gap
$\sim$0.15~eV exactly at the Fermi-surface.  That means 

%%%%%%%%%%%%%%%%%%%%%%%%%%%%%%%%%%%%%%%%%%%%%%%%%%%%%%%%%%%%%
%%%%%%%%%%%%%%%%%%%%%%%%%%%%%%%%%%%%%%%%%%%%%%%%%%%%%%%%%%%%%

\noindent that CrO$_2$ in an antiferromagnetic phase would have been
an insulator.  Thus, the metallicity and ferromagnetism in CrO$_2$ do
indeed support each other, which is in full agreement with the double
exchange picture.

The reason for this dualistic behaviour of the t$_{2g}$ states i.e
some localized and others dispersive, can be found in the distortion
from an ideal rutile-type crystal structure \cite{ideal}.  Our
calculations show that in the ideal structure with the same length of
all 12 O-O bonds in a perfect oxygen octahedron, all the t$_{2g}$
bands would have the same width.  In the real distorted structure the
edge sharing nearest Cr neighbors (Cr1 and Cr2 in Fig.\ref{orb}) move
apart decreasing the $xy$ bandwidth and shifting its center of gravity
to low energies in comparison with the centers of gravity of the other
two t$_{2g}$ bands.  And now with the inclusion of Coulomb
interactions, this $xy$ band becomes the first candidate to be fully
occupied.  Just that happened in our LSDA+U calculation.  This picture
has much in common with the band structure, proposed by Goodenough
\cite{Good} on phenomenological grounds.

It is possible to visualize the $d$ orbitals mentioned above (in
particular, occupied orbitals) by plotting the angular distribution of
$d$ electron spin density for these orbitals (Fig.\ref{orb}):

\begin{equation}
\rho(\theta,\phi)=\sum_{m,m'}Q_{m,m'}Y_m(\theta,\phi)Y_{m'}(\theta,\phi)
\end{equation}

\noindent where $Q_{m,m'}=n^{\uparrow}_{m,m'}-n^{\downarrow}_{m,m'}$
is the $d$ spin occupation number for the case of $xy$ orbital or
2$\ast$2 matrix for the case of ($yz+zx$) orbital obtained in our
self-consistent calculation and $Y_m(\theta,\phi)$ are corresponding
spherical harmonics.  Note that ($yz+zx$) orbitals of Cr atoms
hybridize with each other via $\pi$-overlap with O1 and O2 oxygens
(right panel of Fig.\ref{orb}).

Yet another interesting feature is that there are bands of
predominantly O $2p$ character crossing the Fermi level close to the
$\Gamma$-point (Fig.\ref{bands}).  The absence of hybridization of
these with the $d$ states indicates that they have symmetries such
that this mixing is not allowed.  The fact that they extend to the
Fermi level is important.  This means that these states can be used as
electron or hole reservoirs causing a non-integral occupation of the
$d$ bands.  This may be called self-doping.  This again is a reason
for the metallic behaviour in spite of quite large U values.

The situation when the almost pure O $2p$ bands cross the Fermi level
indicates that such materials should be considered as lying in the
small or even negative charge transfer gap region of the ZSA diagram.
The negative charge transfer gap behaviour is not unexpected for
oxides with transition metals in very high oxidation states because of
the increased electron affinity of these ions.  This conclusion can in
principle be checked by the $1s$-$2p$ x-ray absorption on oxygen;
simultaneously certain spin polarization should be transferred to
oxygen which would be very interesting to study experimentally.  In
our calculations the oxygen polarization turned out to be opposite to
the polarization of $d$-ions and equal to -0.09~$\mu_B$ per oxygen.

One more conclusion of the picture obtained is that, as is common for
the double exchange systems, one should expect a negative
magnetoresistance in the paramagnetic region.  Previous results have
shown that the magnetoresistance of CrO$_2$ is rather small
\cite{Chamb_rev,mr}, but these measurements were done only below room
temperature.  It would be interesting to extend these measurements to
the region of T$>$T$_c$=392~K.  As there is no extra "complications"
in CrO$_2$ as compared to CMR manganates (no strong Jahn-Teller
effects, no disorder, no formally different valence ions), this study
would reveal a pure double exchange contribution to magnetoresistance
which, when compared with similar data for manganates, could help to
discriminate between the contribution of different mechanisms in the
latter.

Summarizing, we carried out an extensive band structure calculations
of CrO$_2$ in the LSDA+U approach.  We confirmed the half-metallic
nature of CrO$_2$, albeit with the dip of DOS at the Fermi level.  It
is demonstrated that there exist in CrO$_2$ two groups of
$d$-electrons with significantly different properties:  one of the two
$d$-electrons of Cr$^{4+}$ is essentially localized, and another is
$\pi$-bonded with the oxygen $p$-orbitals and forms a partially filled
narrow band.  The width of this band is of the same order as the
Hund's rule intraatomic exchange interaction.  The resulting picture
may be interpreted as an indication that the ferromagnetism in CrO$_2$
is due to double exchange mechanism.  An extra confirmation of this
conclusion comes from the fact that in an antiferromagnetic phase
CrO$_2$ would have been an insulator.  The strong contribution of the
oxygen $p$-states at the Fermi surface explains why this material is
not a Mott insulator in spite of the large U values and the relatively
narrow $d$-bands and also shows that CrO$_2$ belongs to the class of
compounds with small or even negative charge-transfer gap leading to a
self-doping.  Several experiments are suggested which could check the
proposed picture and could help to isolate the double exchange
contribution to magnetoresistance.

This investigation was supported by the Russian Foundation for
Fundamental Investigations (RFFI grant 96-02-16167) and by the
Netherlands Organization for Fundamental Research on Matter (FOM),
with financial support by the Netherlands Organization for the advance
of Pure Science (NWO).

\end{multicols}


\begin{thebibliography}{99}

\bibitem{Zener} C. Zener, Phys. Rev. {\bf 82}, 403 (1951).

\bibitem{KhS_rev} D.I.  Khomskii, and G.A.  Sawatzky, Solid State
Comm.  {\bf 102}, 87 (1997).

\bibitem{LSDA+U} V.I. Anisimov, J. Zaanen, and O.K. Andersen, Phys.
Rev. {\bf B44}, 943 (1991). A.I. Lichtenstein, J. Zaanen, and 
V.I. Anisimov, Phys. Rev. {\bf B52}, R5467 (1995).

\bibitem{neggap} T. Mizokawa, H. Namatame, A. Fujimori, K. Akeyama,
H. Kondoh, H. Kuroda, and N. Kosugi, Phys. Rev. Lett. {\bf 67}, 1638
(1991).

\bibitem{ZSA} J. Zaanen, G.A. Sawatzky, and J.W. Allen, Phys. Rev.
Lett. {\bf 55}, 418 (1985).

\bibitem{LJP} D.  Khomskii, to be published in the R.  Dagys memorial
issue of the Lithuanian Journal of Physics.

\bibitem{Chamb_rev} B.L. Chamberland, CRC Crit. Rev. Solid State Sci.,
{\bf 7}, 1 (1977).

\bibitem{Schwarz} K. Schwarz, J. Phys. {\bf F16}, L211 (1986).

\bibitem{Kamper} K.P.  K\"amper, W.  Schmitt, G.  G\"untherodt, R.J.
Gambino, and R.  Ruf, Phys.  Rev.  Lett.  {\bf 59}, 2788 (1987).

\bibitem{cryststr} B.J. Thamer, R.M. Douglass, and E. Staritzky,
J. Am. Chem. Soc. {\bf 79}, 547 (1957).

\bibitem{lmto} O.K. Andersen, Phys. Rev. {\bf B12}, 3060 (1975).

\bibitem{calcU} W.E.  Pickett, S.E.  Erwin, and E.C.  Ethridge,
preprint cond-mat/9611225 v3.

\bibitem{Fujimori} T.  Tsujioka, T.  Mizokawa, A.  Fujimori, M.
Nohara, H.  Takagi, K.  Yamaura, and Y.  Takano, unpublished.

\bibitem{ideal} For detailed description of ideal rutile-like crystal
structure see, e.g., K.M.  Glassford, and J.R.  Chelikowsky, Phys.
Rev.  {\bf B46}, 1284 (1992).

\bibitem{Good} J.B.  Goodenough, in {\em Progress of Solid State
Chemistry}, edited by H.  Reiss, Vol.~{\bf 5}, 145 (Pergamon, Oxford,
1971).

\bibitem{mr} L. Ranno, A. Barry, and J.M.D. Coey, preprint

\end{thebibliography}
\end{document}